\documentclass[fleqn,twoside]{article}
\usepackage{espcrc2}
\usepackage{graphicx}

\newcommand {\mm}       {\ensuremath}%
\newcommand {\mr}       {\mathrm}%
\newcommand {\CP} {$C\!P$}
\newcommand {\ra}        {\mm{\rightarrow}}%
\newcommand {\vu}    [2] {\mm{#1\:\mr{#2}}}                 
\newcommand {\etal}   {et~al.}
\newcommand {\ie}     {i.e.~}
\newcommand {\vvu}   [3] {\mm{#1\!=\vu{#2}{#3}}}            

\newcommand {\He}       {\mm{\mr{He}}}
\newcommand {\aHe}      {\mm{\overline{\mr{\!He\!}}}}

\newcommand {\MeV}  {\ifmmode{\mr{Me\kern -0.1em V}}%
                    \else\textrm{Me\kern -0.1em V}\fi}%
\newcommand {\GeV}  {\ifmmode{\mr{Ge\kern -0.1em V}}%
                    \else\textrm{Ge\kern -0.1em V}\fi}%
\newcommand {\R}        {\mm{R}}

\newcommand{\N}{\ensuremath{Z}}
\newcommand{\aN}{\ensuremath{{\,\overline{\!Z\!}}\,}}
\newcommand{\bfc}{\begin{figure}[htb]\centering}
\newcommand{\efc}{\end{figure}}
\newcommand{\epsinc}[3]{\includegraphics[width=#2pc]{#1}\vspace{-5mm}\caption{{\it #3}}\label{#1}}
\newcommand{\beps}[3]{\bfc \epsinc{#2}{#1}{#3}\efc}

\title{Search for Antimatter with the AMS Cosmic Ray Detector}

\author{Markus Cristinziani\address{%
D.P.N.C., Universit\'e de Gen\`eve, CH-1211 Gen\`eve 4,
Switzerland\\Now at: {\it Stanford Linear Accelerator Center, Stanford, CA 94309, USA}}%
\thanks{markus@slac.stanford.edu}}

\begin{document}

\begin{abstract}
Antimatter search results of the Alpha Magnetic Spectrometer (AMS) detector are presented.
About $10^8$ triggers were collected in the 1998 precursor flight onboard space shuttle {\it Discovery}.
This ten day mission exposed the detector on a 51.7$^\circ$ orbit at an altitude 
around $350 \mr{km}$. 
Identification of charged cosmic rays is achieved by multiple energy loss and
time-of-flight measurements. Bending inside the $0.15 \mr{T}$ magnetic volume 
yields a measurement of the absolute value of the particle's rigidity. The supplemental
knowledge of the sense of traversal identifies the sign of the charge.
In the rigidity range \mbox{$1<R<140$ GV} no antinucleus at any rigidity was detected,
while \mbox{$2.86 \times 10^6$} helium and \mbox{$1.65 \times 10^5$} heavy nuclei 
were precisely measured.
Hence, upper limits on the flux ratio $\aN/\N$ are given. Different prior assumptions on the
antimatter spectrum are considered and corresponding limits are given.
\end{abstract}

\maketitle

\section{Introduction}

The Dirac theory of elementary interactions states that
to each particle corresponds an antiparticle with all additive quantum
numbers inverted in sign, and of each reaction involving particles,
the symmetric can occur.
An experimental confirmation was provided by the discovery of the positron
in cosmic rays \cite{and33} and later by the production of antiprotons
at accelerators \cite{cha55}.
However we do not observed this symmetry in our macroscopic surrounding
which is composed of photons and matter particles. Indeed, there is no
undisputed experimental evidence up to now which proves the presence of
antimatter in our Universe.

Based on the evidence that in weak interactions parity ($P$) is not conserved
\cite{wu57} nor is the product of charge conjugation and parity (\CP) in the
very specific case of the neutral $K$ system \cite{chr64}, Sakharov \cite{sak67}
pointed out that three ingredients are necessary for a baryon
symmetric Universe to evolve to an asymmetric one:
\begin{itemize}
\item violation of the baryon number ($B$);\\[-3.5ex]
\item violation of $C$ and \CP;\\[-3.5ex]
\item departure from thermal equilibrium.
\end{itemize}

There is no compelling reason for the baryon number to be strictly conserved
and there is no evidence of a force associated with baryonic charge.
In fact, GUT theories predict $B$ violating interactions.
There are several experiments which are testing $B$ nonconservation
by searching for neutron-antineutron oscillations
or proton decays, none of them having detected a significant signal so far.
The present world limit on the proton lifetime, determined by the partial width
of the decay $p \ra e^+ \pi$, is
$\tau_p > 1.6 \times \vu{10^{33}}{yrs}$ \cite{pdg00}.
The strength of the observed \CP\ violation is far too small to account for the baryon asymmetry of the Universe.

Antimatter does not exist on Earth in macroscopic amounts: it would have
been annihilated releasing tremendous amounts of energy.
The solar wind, the constant flux of charged particles emitted by
the Sun and propagated throughout the solar system, allows us to
exclude antimatter planets, which otherwise would appear as very
bright $\gamma$-ray emitters.
Photons emitted by other stars do not probe directly the sign of the baryon
number of the object they are emitted from.
Fortunately cosmic rays do and we can therefore expect to learn about the
baryon content of the Galaxy and the Universe by studying their composition.

\section{Summary of previous measurements}

Balloon-borne experiments started to search for signatures of 
antinuclei in their instruments since 40 years. 
A summary of the most stringent limits 
on the antihelium flux as a function of the observed rigidity range can
be found in reference \cite{elba}.
%
The BESS collaboration deploy their detector, whose central part is
made out of a drift chamber enclosed by a thin superconducting magnet, 
on balloon flights from a high latitude location on an annual basis 
since 1993. 
The latest limit on the antihelium-to-helium flux ratio reported by the 
BESS collaboration is $7\times10^{-7}$ \cite{sas01} up to a maximum 
rigidity of \vu{14}{GV}.


A comprehensive compilation of limits on the antimatter-to-matter flux
ratio is shown in fig.~\ref{fig1.eps}. 

\beps{14}{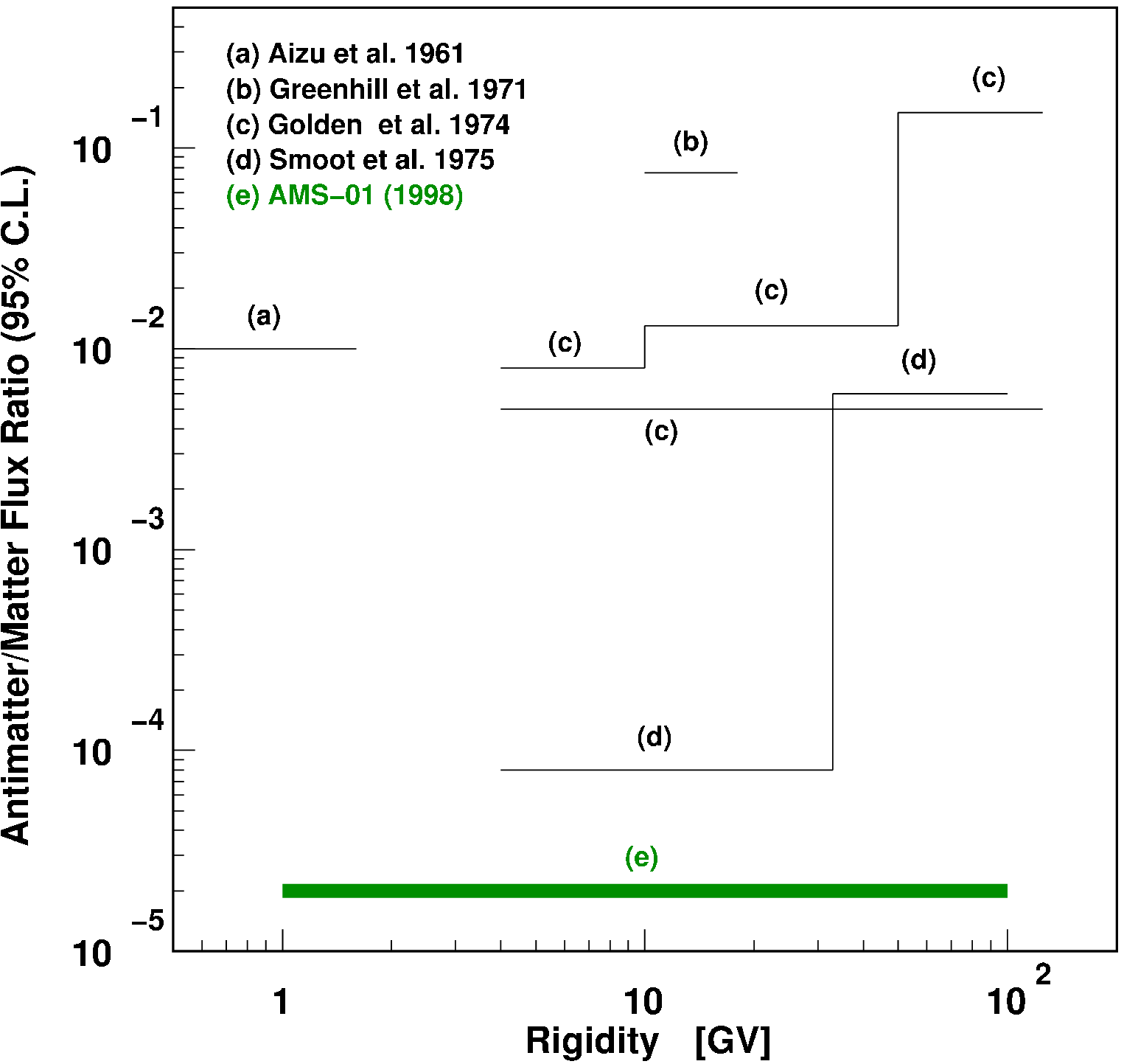}{Compilation of limits on the antimatter-to-matter flux ratio. The AMS-01
measurement provides an improvement both in sensitivity and in rigidity range.}

A stack of photographic emulsions
was used in 1961 \cite{aiz2} to detect annihilation topologies with a 
maximum detectable energy of \vvu{E_\mr{max}}{700}{\MeV/n}.
Greenhill and collaborators \cite{gre}
employed a gas \v{C}erenkov detector and scintillators for the $|\beta|$
and $dE/dx$ determination in 1971. The traversal direction could not be
directly measured but was derived by considerations on the different
dependence of the cutoff rigidity for negative and positive charged
particles for a given orientation of the instrument with respect to zenith.
Because of the higher \v{C}erenkov threshold it was possible to exclude the presence
of antimatter up to energy values of \vvu{E}{9}{\GeV/n}.
The first magnetic spectrograph with spark chambers and emulsion
plates recorded a relatively small amount
of events, while Golden \etal\ in 1974 \cite{gol1} and Smoot \etal\ in 1975 \cite{smo}
used a superconducting magnet with a bending power of \vvu{BL}{0.43}{Tm}.
With an acceptance of \vu{0.066}{m^2\mr{sr}} a total of $\sim 10^4$
events were collected up to rigidity values
of \vvu{R}{100}{GV}. This measurement yielded the most stringent
limit for the presence of antinuclei ($Z<-2$) before AMS.

\section{The AMS apparatus}

\beps{14}{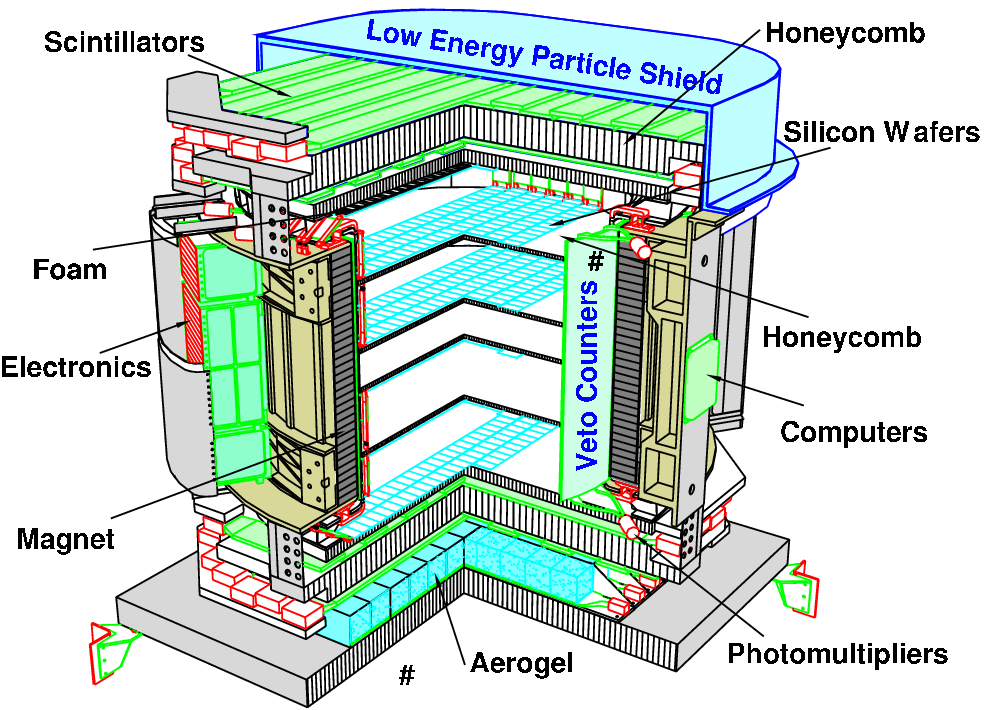}{The AMS-01 detector as flown on space shuttle Discovery in 1998.}

The Alpha Magnetic Spectrometer (AMS)~\cite{ahl94} is scheduled
for a high energy physics program on the International
Space Station.
The primary goal of the AMS experiment
is the search for antimatter in cosmic rays.
The unambiguous signature which is looked for are nuclei with $|Z| > 1$.
The physical quantities which are measured by the AMS detector include
the particle charge and mass for its identification, its energy or equivalently
its rigidity, and the sense and direction of traversal.

The AMS-01 \cite{vie98} version of the Alpha Magnetic Spectrometer is schematically shown in
fig.~\ref{fig2.eps}. 
It was flown on the
space shuttle {\it Discovery} on flight STS--91 in June 1998 for a ten days test flight.
The cylindrical permanent magnet encloses a multilayer
silicon tracker which measures the trajectory of charged particles traversing
the volume. Four scintillator planes and an Aerogel Threshold \v{C}erenkov
counter complete the detector by measuring the particle velocity. The
energy loss is recorded by the tracker and the scintillators.
In order to reject particles outside the magnet aperture the inner wall of the magnet
is covered by an anticoincidence counter. To minimize dead time a
``Low Energy Particle Shield'' absorbs low energy particles above the scintillator
planes.
For particles arriving from above,
the amount of material at normal incidence
was $1.5\,\mathrm{g/cm}^2$ in front of the TOF system,
and $3.5\,\mathrm{g/cm}^2$ in front of the tracker.

\section{Search for antinuclei}

A detailed description of the search for antihelium is given in reference
\cite{alc}.\\
The goal of the analysis is to find a small amount of antimatter in a large
background of matter. A total of 270905 events are identified as
particles traversing the detector with $|Z|>2$, out of which about
6\% are initially reconstructed as antimatter candidates:
\begin{itemize}
\item Number of events with $Z>2$: 255321\\[-5ex]
\item Number of events with $Z<-2$: 15584.
\end{itemize}

\beps{14}{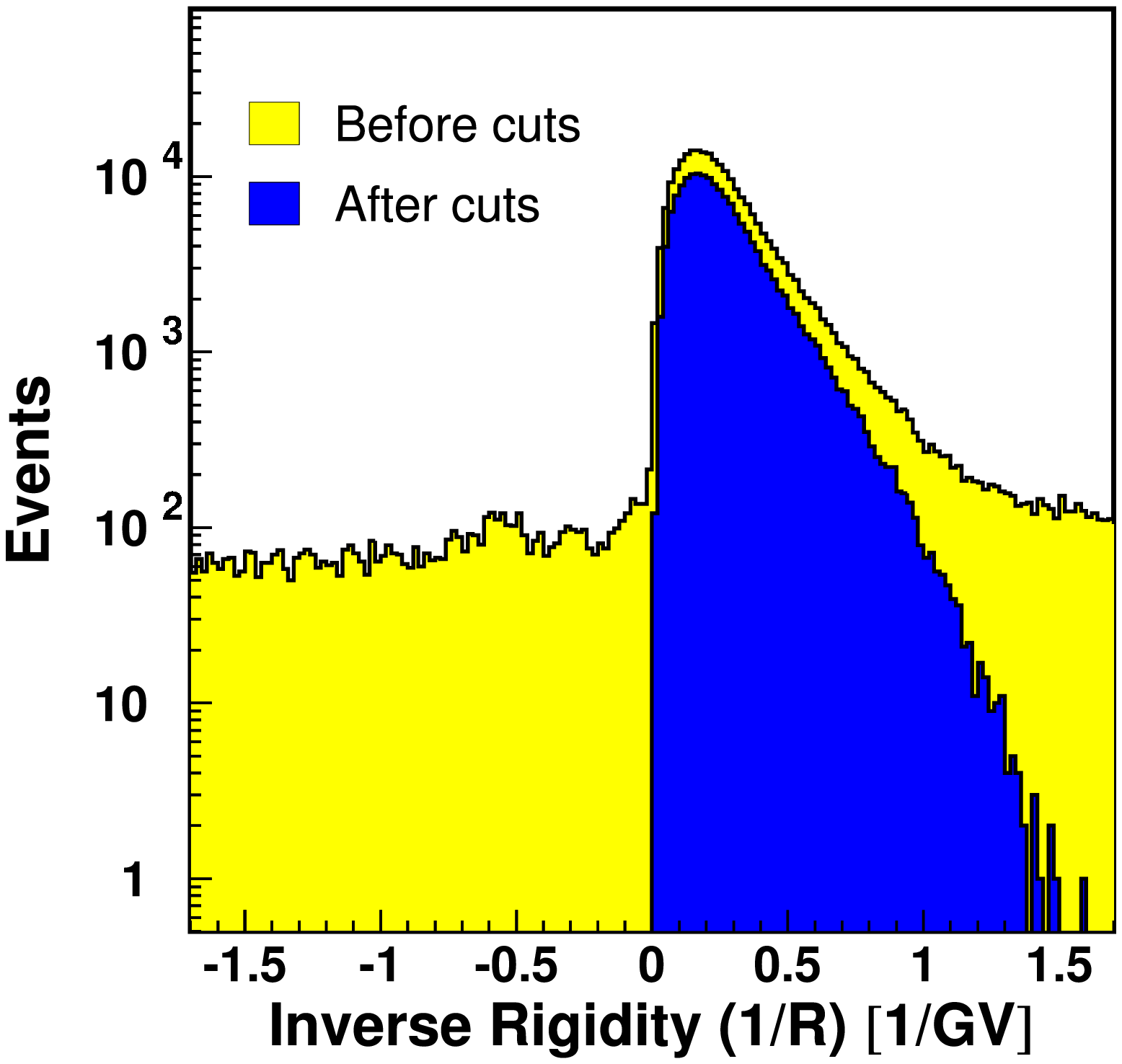}{Event distribution as a function of the inverse rigidity. The event quality cuts
remove all $R<0$ candidates.}

In fig.~\ref{fig3.eps} rigidity distributions are shown before and after
all selection cuts.
The measured parameter is the track's sagitta, \ie the deflection,
which is proportional to the inverse rigidity $1/R$.
The candidate antimatter events are uniformly distributed as a function of
$1/R$, which immediately suggests that they predominantly arise from
misreconstruction.
The event selected with the lowest rigidity has
a value \vvu{1/R}{1.59}{GV^{-1}} or \vvu{R}{630}{MV}.
At lower values a background level of $\sim 100$ events/bin is reached.


A matter nucleus may in principle mimic an antimatter
nucleus flying in opposite direction if the time-of-flight is
determined with the wrong sign.
Due to the good time resolution of the TOF system the $\beta$ measurement
is sufficiently accurate not to mix the two particle populations coming
from the top and from the bottom of the detector. A misidentification
is therefore excluded. 
%

The dynamic range of the silicon tracker allows a good separation of nuclei
at least up to oxygen ($\mr{Z}=8$). Up to six measurements are available 
for the same particle. The energy lost inside each silicon detector is 
measured twice (``S'' and ``K'' side).
Fig.~\ref{fig4.eps} shows the obtained separation capability.

\beps{16.13}{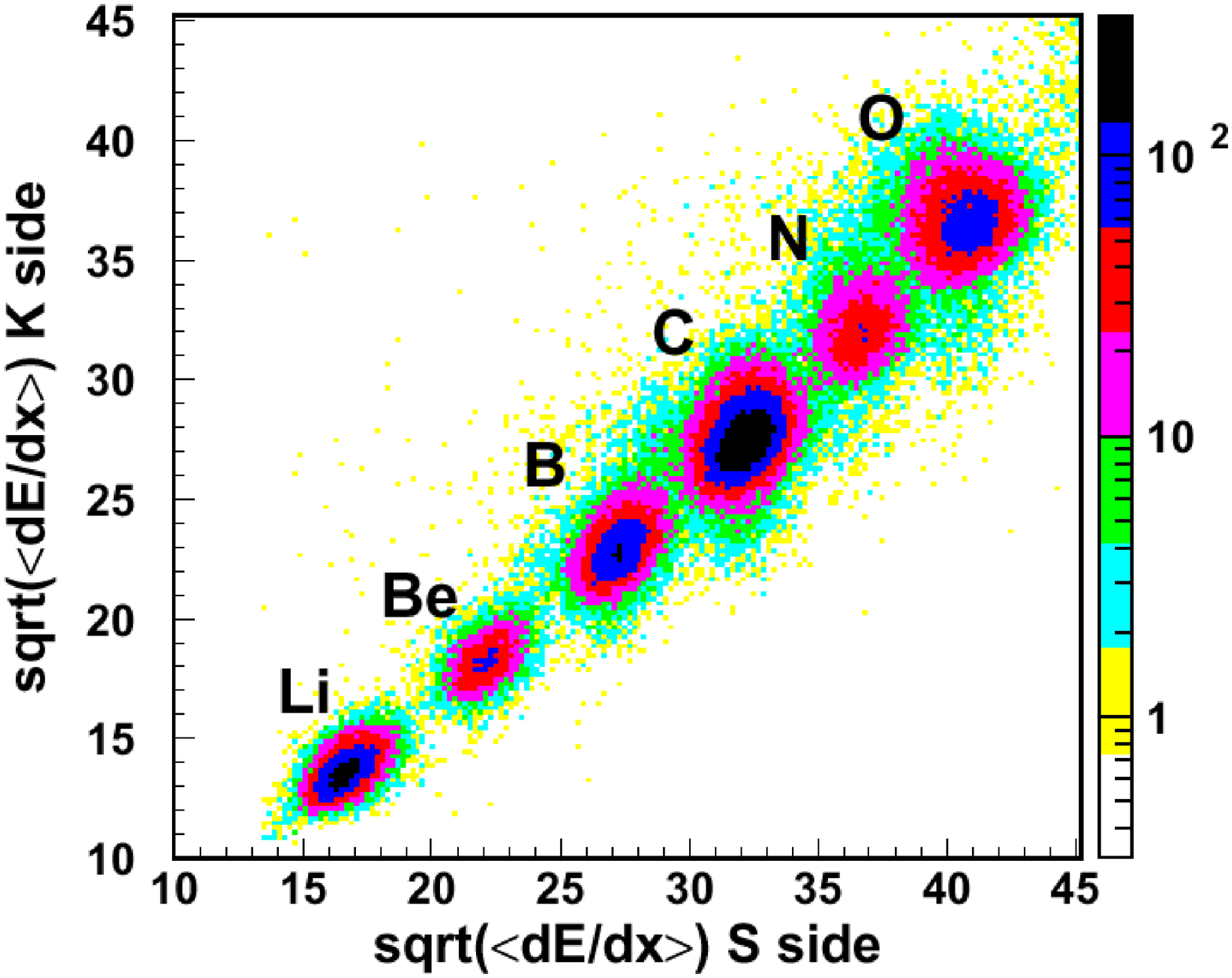}{Combined charge measurement by energy loss recorded for S side and
K side clusters.  Peaks are represented on a logarithmic scale.}

A set of cuts has been developed~\cite{mythesis} to ensure the selection 
of clean events. Special care has to be taken to eliminate events which 
have a poor determination of the track parameters. For example, the rigidity 
estimations obtained using the upper and lower three tracking points are 
required to agree.

The last cut is applied to ensure the overall consistence of the
velocity, rigidity and charge measurements.
Fig.~\ref{fig5.eps} shows the distribution of 1/$\beta$ versus rigidity together
with the cuts applied for $|Z|>2$ events.
True $|Z|>2$ events should be concentrated
along the line $1/\beta= \sqrt{1+(A M_n/Z R)^2}$, where A is the atomic
number and $M_n$ the nucleon mass. Measurement errors in $\beta$, $Z$ or \R\
cause scattering around this line.  The shown cuts reject all remaining
$Z<-2$ antimatter candidates and keep nearly all the $Z>2$ events.

\beps{16.68}{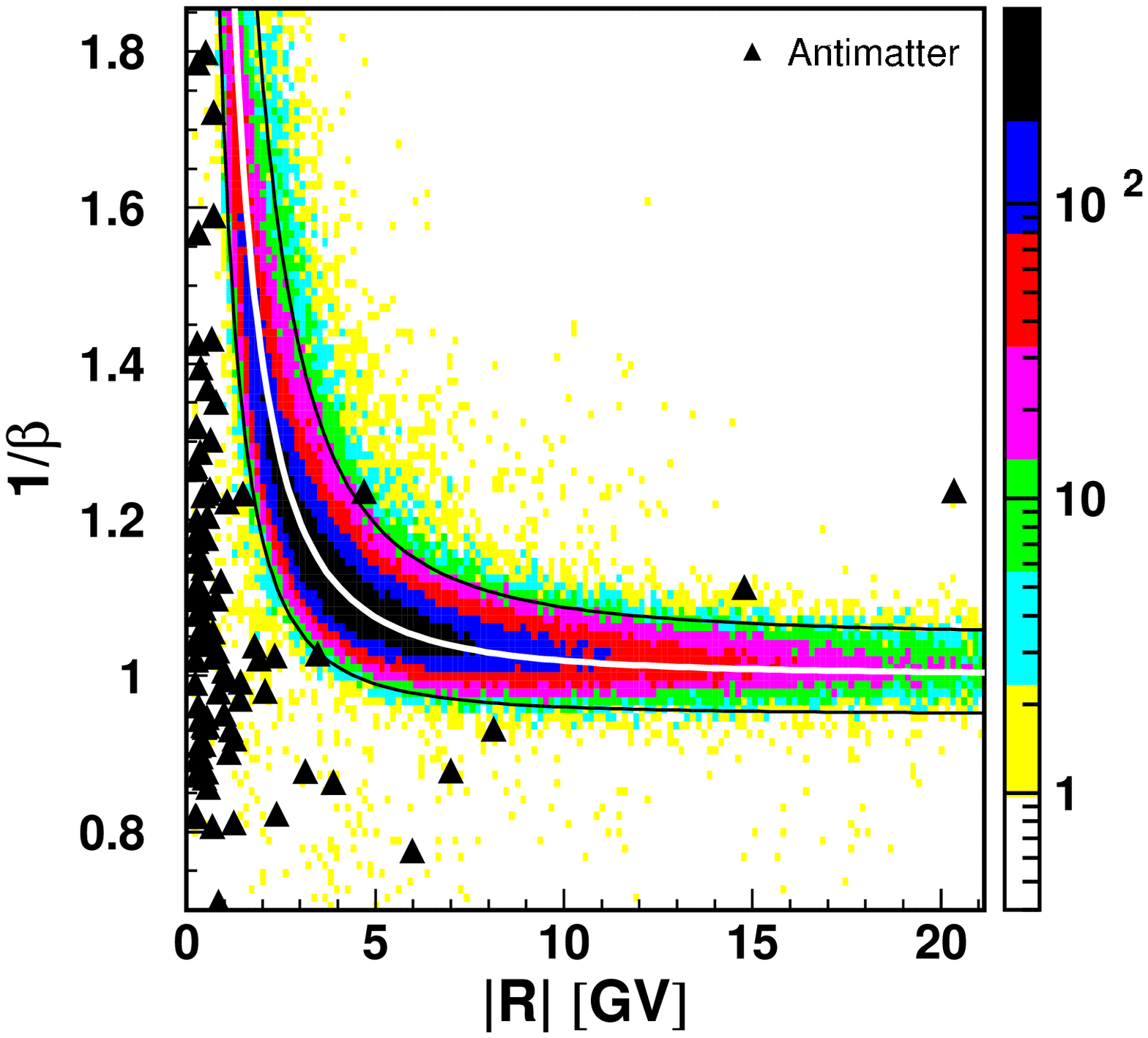}{Time-of-flight vs. absolute rigidity~\cite{mxc01}.
Nuclei with $(mc^2/Ze)^2 \sim \vu{4}{GV^2}$ should
lie along the white line. Well reconstructed events are selected within
the black lines. All the antimatter candidates (triangles) are outside
this region.}

Corrections to the measured spectrum take into account the rigidity 
resolution function, the detector livetime and rigidity cutoff dependence
on the latitude, and the different cross section of nuclei and antinuclei
hitting the detector material.

\section{Limits on the antimatter-to-matter flux ratio}
\renewcommand{\He}       {\mm{\!Z\!}}
\renewcommand{\aHe}      {\mm{\overline{\!Z\!}}}

Since no antimatter nucleus was found at any 
rigidity, one can provide an upper limit on the flux ratio of
antimatter to matter $N_{\aHe}/N_{\He}$
which is evaluated by summing up the contents of all the bins
\[
  \left[
  \frac{N_{\aHe}}{N_{\He}}\right]=
  \frac{\sum{N_{\aHe}(\R_i)}}{\sum{N_{\He}(\R_i)}}.
\]
It should be noted that the experimental result of detecting no
antimatter nucleus corresponds to $\sum{N^\prime_{\aHe}(\R_i)}=0$,
where $N^\prime$ denotes the {\it measured} spectrum.

Following the unified approach of confidence belts construction 
proposed in \cite{fel98} which ensures correct coverage avoiding 
unphysical confidence intervals and is based on classical statistics,
we can put $\sum{N^\prime_{\aHe}(\R_i)} < 3.09$ at a $95\%$ confidence
level.

Three different priors can be assumed for the unknown
antimatter distribution,
$N_{\aHe}(\R_i)$:
\begin{itemize}
\item the same spectrum as matter, $N_{\aHe}(\R_i)/N_{\He}(\R_i) = \mr{const}$;\\[-4ex]
\item a flat spectrum, $N_{\aHe}(\R_i)=\mr{const}$;\\[-4ex]
\item no a priori spectrum (worst case), $N_{\aHe}(\R_j)= N_{\aHe}$, 
where j is the bin with the lowest efficiency and $N_{\aHe}(\R_i)=0$ otherwise.
\end{itemize}

Limits for the three approaches are calculated as follows (see reference \cite{elba} 
for details):

\[
  \:\:\:\:\:\:\:\:\:\:\:\:\:\:\:\left[\frac{N_{\aHe}}{N_{\He}}\right]_\mr{same}  < 
  \frac{3.09}{\sum{N^\prime_{\He}(\R_i)\frac{\epsilon_{\aHe}(\R_i)}
  {\epsilon_{\He}(\R_i)}}},\\
\]
\[
  \:\:\:\:\:\:\:\:\:\:\:\:\:\:\:\left[\frac{N_{\aHe}}{N_{\He}}\right]_\mr{unif}  < 
  \frac{3.09/\!<\!\epsilon_{\aHe}\!>}
  {\sum{N^\prime_{\He}(\R_i)/\epsilon_{\He}(\R_i)}},\\
\]
\[
  \:\:\:\:\:\:\:\:\:\:\:\:\:\:\:\left[
  \frac{N_{\aHe}}{N_{\He}}\right]_\mr{cons}  <
  \frac{3.09/\epsilon_{\min}}
       {\sum{N^\prime_{\He}(\R_i)/\epsilon_{\He}(\R_i)}}.
\]

where $\epsilon_{\He}(\R_i)$ and $\epsilon_{\aHe}(\R_i)$ are the detector efficiencies 
for matter and antimatter, the {\it average} efficiency is defined as
\mbox{$<\!\epsilon_{\aHe}\!> \equiv (\sum \epsilon_{\He}(\R_i))/n$}, and
$\epsilon_{\min}$ is the worst efficiency in a given rigidity interval.

Under the same spectrum assumption
$\left[N_{\aHe}/N_{\He}\right]_\mr{same}< 2.00 \times 10^{-5}$ for all nuclei.
This result is compared to previous measurements in 
fig.~\ref{fig1.eps}. A more stringent limit could be given due
to the amount of events collected and the good rigidity resolution.
The rigidity range extends from \vu{1}{GV} to \vu{100}{GV}.

In the second case, the upper limit on the anticarbon/carbon ratio, for example,
is $6.55 \times 10^{-5}$ in the rigidity range \vu{1}{GV} to
\vu{10}{GV} and $1.46 \times 10^{-4}$ in the range \vu{1}{GV} to \vu{50}{GV}.

\beps{14.41}{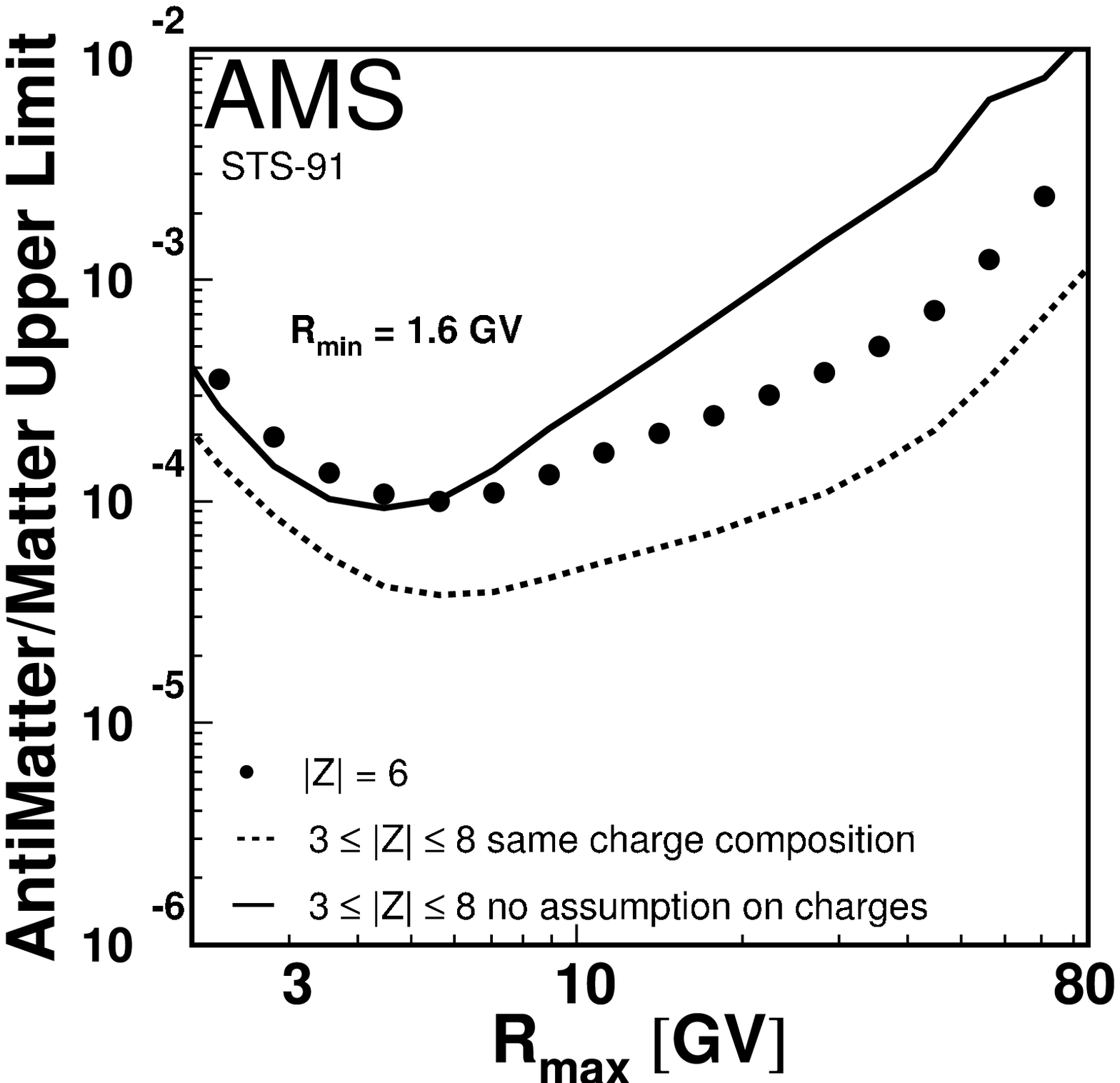}{Upper limits on the antimatter-to-matter flux ratio
under the conservative approach.
Integrating over the rigidity range
$[\R_{\min}=\vu{1.6}{GV};\R_{\max}]$, the limit curves are shown as a
function of the maximal rigidity $\R_{\max}$.}

The upper limit determined under the most conservative assumption
is shown in fig.~\ref{fig6.eps}
for all matter/antimatter nuclei ($3 \le |Z| \le 8$) and for 
carbon/anticarbon.
Starting from a minimum rigidity of $\R_{\min}=\vu{1.6}{GV}$ the
upper limit is shown as a function of the maximum rigidity.
With increasing $\R_{\max}$ the limits decrease as a consequence of the
increasing statistics. The rise above $\R_{\max} \sim \vu{4}{GV}$ is
due to the detection efficiency which gets worse with increasing 
rigidity. The corresponding result for antihelium can be found in~\cite{alc}.


With the AMS-01 test flight a never before obtained sensitivity for antimatter searches
has been reached. The AMS-02 version of the detector which will include a superconducting
magnet and will be installed on the International Space Station is foreseen to further
improve this sensitivity up to TV rigidities~\cite{jorge}.

\end{document}